\documentclass[amsmath,amssymb,aps,superscriptaddress, twocolumn, print]{revtex4-2}

\usepackage[english]{babel}
\usepackage[utf8x]{inputenc}
\usepackage{amsmath}
\usepackage{graphicx}
\usepackage{comment}
\usepackage[colorinlistoftodos]{todonotes}
\usepackage{xcolor,soul}
\definecolor{cerclegreen}{HTML}{139F46}

\makeatletter
\renewcommand{\fnum@figure}{FIG. \thefigure}
\renewcommand{\fnum@table}{TABLE \thetable}
\makeatother

\begin{document}

\title{Commensurate vortex core switching in magnetic nanodisks at Gigahertz frequencies}

\author{Pieter Gypens}
\affiliation{Dept. of Solid State Sciences, Ghent University, Belgium}

\author{Jonathan Leliaert}
\affiliation{Dept. of Solid State Sciences, Ghent University, Belgium}

\author{Gisela Sch\"utz}
\affiliation{Max Planck Institute for Intelligent Systems, Stuttgart, Germany}

\author{Bartel \surname{Van Waeyenberge}}
\email{bartel.vanwaeyenberge@ugent.be}
\affiliation{Dept. of Solid State Sciences, Ghent University, Belgium}

\begin{abstract}
The development of future spintronic applications requires a thorough and fundamental understanding of the magnetisation dynamics. Of particular interest are magnetic nanodisks, in which the vortex state emerges as a stable spin configuration. Here, we focus on how the vortex core polarisation can be reversed periodically by an oscillating magnetic field, applied perpendicularly to the disk's surface. By means of micromagnetic simulations, we demonstrate the presence of several subharmonic switching modes, i.e., the commensurate ratio between the switching frequency of the core and the driving frequency. The underlying mechanism of this periodic behaviour depends on the disk thickness. For thin disks, the core switches periodically due to resonant excitation of radial spin wave modes, while it is due to the breathing mode in the case of thick disks. However, overlap of both modes impedes periodic vortex core switching. For thin disks, the threshold field amplitude required for periodic switching can be lowered to about 30~mT by increasing the disk diameter. For thick disks, in contrast, the minimal field is largely unaffected by the disk diameter, as only the energy density of a central region around the vortex core is relevant to excite the breathing mode. Our results contribute to the understanding of the switching mechanisms in magnetic nanodisks, which are of technological interest due to their potential in non-volatile memory devices.
\end{abstract}

\maketitle

\section{Introduction}
In recent years, several applications have been proposed whose operation relies on the magnetisation dynamics in nanostructures, ranging from non-volatile memory devices~\cite{ZHA-04, BOH-08, PAR-08, SAM-13}, to logic devices~\cite{WAN-05, ALL-05, IMR-06, ZHA-15, LUO-20}, over spin-wave based communication and information processing devices~\cite{DIE-19, HAN-19, YU-20}. It is therefore of fundamental importance to understand the magnetisation dynamics in such structures, e.g., magnetic nanodisks in which the vortex state appears due to the competition between the short-range exchange interaction and the long-range magnetostatic interaction. In such a disk, the magnetisation curls in-plane in favour of the magnetostatic interaction, except at the centre where the exchange interaction starts to dominate. This central region, called the vortex core, has a radius of about 10~nm and has an out-of-plane magnetisation~\cite{MIL-02a, WAC-02}. The vortex state is fully characterised by two quantities: the {\it circulation} which refers to the (counter)clockwise curl of the in-plane magnetisation, and the {\it polarisation} of the vortex core which is equal to $p = −1$ for $m_z<0$ and $p = +1$ for $m_z>0$.\\

Mechanisms to reverse the circulation and core polarisation in a reliable, fast, and energetically efficient manner have been studied because of their importance in the development of vortex-based memory~\cite{BOH-08, GEN-17} and logic~\cite{JUN-12, KUM-14} devices. In order to change the circulation, the original vortex core is expelled from the disk by applying an in-plane magnetic field, after which a new core can nucleate at the border~\cite{GAI-08, ANT-09, UHL-13}. The circulation of the newly formed core can be controlled if the rotational symmetry of the disk is broken, e.g., via a thickness gradient~\cite{UHL-13}.\\ 

The vast majority of techniques to reverse the core polarisation are based on the resonant excitation of an eigenmode, as this significantly reduces the required excitation power, compared to a static out-of-plane field~\cite{OKU-02, THI-03, WAN-12b}. Examples of such eigenmodes are the sub-GHz gyration mode~\cite{VAN-06, CUR-08, WEI-09b} and the multi-GHz spin wave modes, including azimuthal~\cite{KAM-11} and radial~\cite{YOO-12, DON-14} modes.\\

The gyration mode is related to the spiralling motion of the vortex core around the centre of the disk which takes place when the core has been displaced from the centre. Methods to excite the gyration mode make use of linear oscillating~\cite{VAN-06}, rotating~\cite{CUR-08}, or pulsed~\cite{WEI-09b} in-plane magnetic fields. The switching is mediated by the creation of a vortex-antivortex pair with polarisation opposite to the original vortex. The antivortex subsequently annihilates with the original vortex, such that only the vortex of opposite polarisation remains, thus completing the reversal~\cite{VAN-06}. This annihilation process involves the injection of a Bloch point, a topological singularity with vanishing magnetisation at its centre~\cite{FEL-65, DOR-68}. \\ 

The spin wave modes are characterised by the radial and azimuthal profile of the out-of-plane oscillations of the magnetisation. Excitation of azimuthal spin waves can be achieved with an oscillating in-plane magnetic field when the rotation sense matches the azimuthal mode number~\cite{KAM-11}. Similar to the gyration mode, the switching of the core polarisation is mediated by the creation of a vortex-antivortex pair. Radial spin wave modes, in contrast, are excited by an oscillating out-of-plane field~\cite{YOO-12, DON-14}. In this case, the core reverses directly due to the injection of a Bloch point. This Bloch point is created at the surface and travels through the disk towards the other surface where it will disappear, effectively changing the polarisation of the vortex core.\\

Besides classifying vortex core switching via the excited mode, a distinction can be made based on the selectivity:  whether the external stimulus only allows for reversals from one state to the other state (e.g., $p=-1 \to p=+1$)~\cite{CUR-08, KAM-11}, or generates switching regardless of the initial core polarisation, such that back and forth toggling between $p=-1$ and $p=+1$ becomes possible~\cite{VAN-06, YOO-12, DON-14}. In this work, we focus on a special case of the latter, where the core reversals occur commensurately to the external stimulus. In order to induce this periodic switching behaviour, we apply an oscillating magnetic field perpendicular to the disk's surface, such that the lateral symmetry is not broken and the core stays at the center of the disk, as opposed to Refs.~\cite{PET-12, YOO-20} where periodic vortex core switching has been observed in a magnetic nanocontact system during self-sustained vortex gyration. Recently, it has been shown that regular switching can be achieved with an alternating out-of-plane field of a few mT, albeit in \emph{curved} magnetic nanodisks which are hard to realise experimentally~\cite{MA-20}.\\ 

Here, by varying the frequency and amplitude of the field in micromagnetic simulations, we obtain switching diagrams which include regions of periodic and aperiodic switching and regions without switching. Specifically, we demonstrate the presence of several subharmonic switching modes in which the periodic behaviour originates either in the resonant excitation of radial spin wave modes in the case of thin disks ($h<35$~nm) or in the \emph{breathing mode} in the case of thick disks ($h>35$~nm). The latter occurs when the frequency of the applied field matches the rate at which the vortex core compresses and expands, similar to the breathing mode of skyrmions~\cite{KIM-14b}. With the aim to lowering the power consumption in potential applications, we investigate how the threshold field needed to induce periodic switching can be lowered by modifying both the size and the material parameters of the disk.


\section{Methods}
We consider circular disks with a diameter of $d=330$~nm and a thickness of $h=21$~nm, for which the ground state is a vortex configuration, as shown in Fig.~\ref{fig_overview}(a). An oscillating magnetic field is applied perpendicularly to the surface of the disk, i.e., ${\bf B}(t) = B_0 \sin (2 \pi f t) {\bf e}_z$. The switching of the core polarisation is investigated for amplitudes $B_0$ ranging from 20 to 100~mT and frequencies $f$ between 7 and 17~GHz. The switching diagrams are acquired by means of micromagnetic simulations, using the software package MuMax3~\cite{VAN-14a}, which numerically solves the Landau-Lifshitz-Gilbert (LLG) equation. Parameters typical for Permalloy are used for the saturation magnetisation ($M_{{\rm s}}=860$~kA/m), the Gilbert-damping constant ($\alpha=0.02$), and the exchange stiffness ($A=13~$pJ/m). The time step is fixed at 0.1 ps.\\ 

The disks are discretised into cells of $(c_x=3~{\rm nm}) \times (c_y=3~{\rm nm}) \times (c_z=7~{\rm nm})$. It has been verified that the results presented in this work are independent of this larger discretisation in the $z$ direction, i.e., the obtained switching diagrams are independent of whether the simulations grid consists of seven layers with $c_z=3~{\rm nm}$ or three layers with $c_z=7~{\rm nm}$.\\

To check whether the core switches \emph{periodically} between $p=+1$ and $p=-1$, we track the core polarisation\footnote{The core polarisation $p$ is determined by a built-in feature of MuMax3, which returns the out-of-plane magnetisation of the cell with the largest $|m_z|$.} for 20~ns. The first 10~ns are sufficient for the system to exit transient states and the subsequent 10~ns are used to determine the time $t_{{\rm min}}$ and $t_{{\rm max}}$, which are the shortest and longest time interval, respectively, for which the sign of the core polarisation remains the same. In the case of periodic switching, $t_{{\rm min}}$ and $t_{{\rm max}}$ are equal, such that their ratio is $r=1$. However, to account for artefacts caused by, e.g., numerical noise, we use a less strict criterion, categorising periodic switching by $r>0.90$. This corresponds to a difference of about 10~ps between the longest and shortest time (i.e., the equivalent of 100 time steps).

\section{Results}

\subsection{Subharmonic switching modes}
\begin{figure*}
\centering{
    \includegraphics[width=0.99\textwidth]{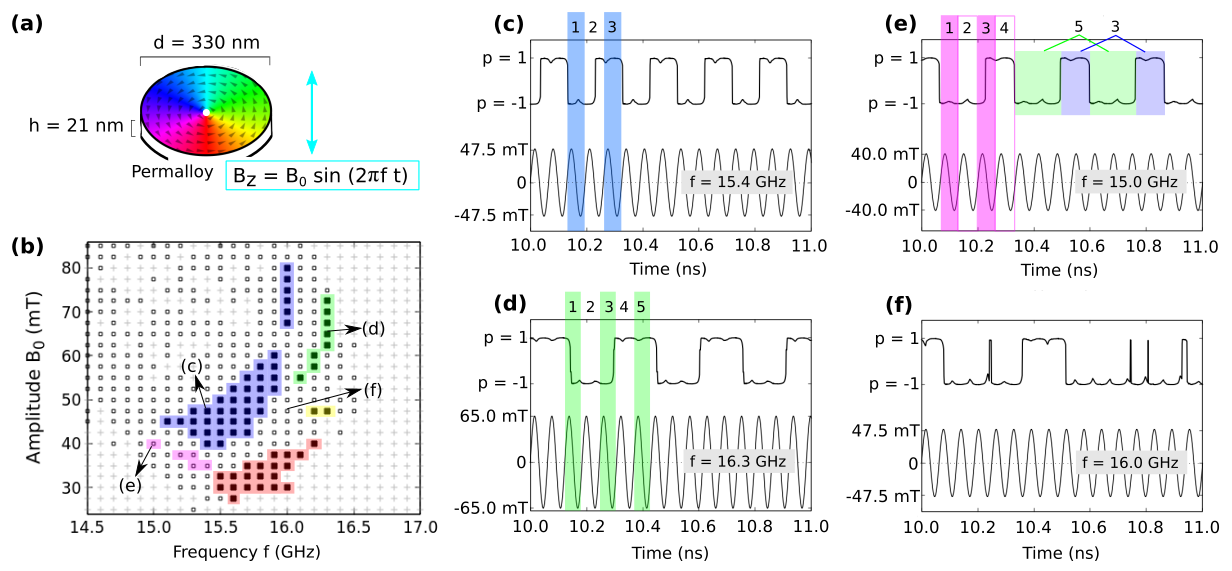}}
    \caption{{\bf (a)} A circular disk is subjected to an oscillating, out-of-plane magnetic field. The color code corresponds to a counterclockwise in-plane magnetisation. At the center, the magnetisation points out-of-plane (white dot). {\bf (b)} The switching diagram shows whether the polarisation of the vortex core reverses (black) or not (gray). Filled squares are used to indicate periodic switching. The colours correspond to different subharmonic switching modes $1/N$, with $N=3$ (blue), $N=5$ (green), $N=7$ (yellow), and $N=9$ (red). The hybrid mode $N^*=4$, which is a superposition of subharmonic $N=3$ and $N=5$, is highlighted in pink. {\bf (c-f)} The applied field and the core polarisation as a function of time for the ($f$,$B$)-points indicated in the switching diagram, showing the difference between the subharmonic switching modes $N=3$ and $N=5$, the hybrid mode $N^*=4$, and an incommensurate mode.}
    \label{fig_overview} 
\end{figure*}

The switching diagram of Fig.~\ref{fig_overview}(b) shows how the regions of periodic switching can be classified according to their subharmonic switching mode $1/N$, with $N$ being an \emph{odd} number. The subharmonic switching mode indicates the ratio between the frequency at which the polarisation of the vortex core reverses and the frequency of the externally applied field. We observe subharmonics for which $N$ is equal to 3 (blue), 5 (green), 7 (yellow), and 9 (red). In Figs.~\ref{fig_overview}(c) and (d), the applied field and the core polarisation are plotted as a function of time for the subharmonics 3 and 5.\\

\begin{figure*}
\centering{
    \includegraphics[width=0.75\textwidth]{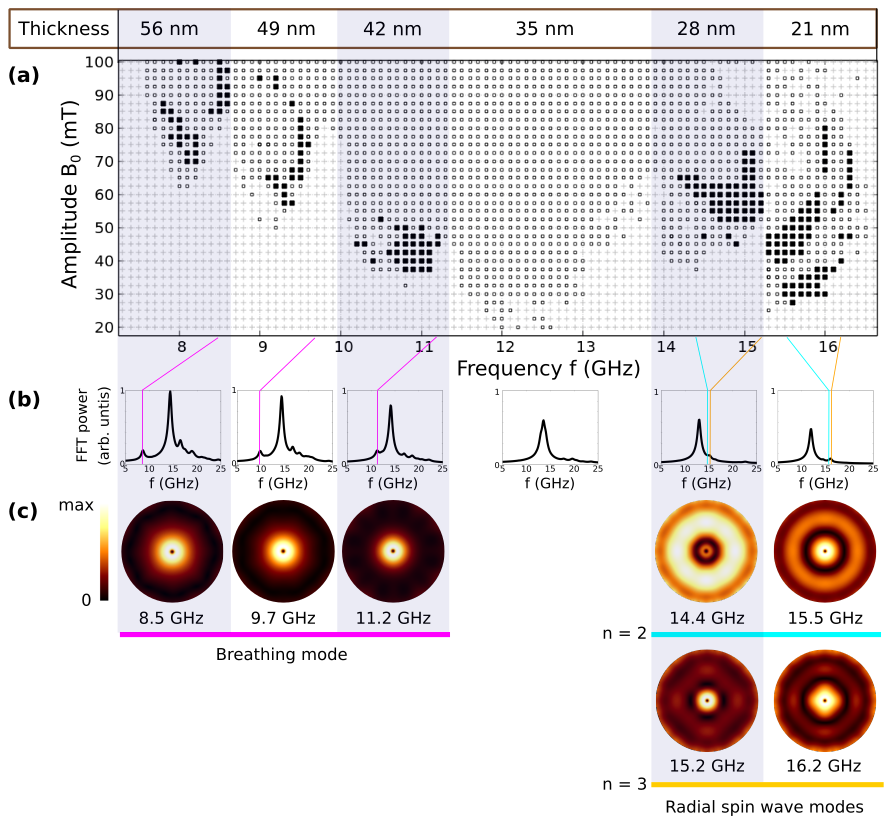}}
    \caption{{\bf (a)} Switching diagrams for different disk thicknesses. The frequencies which give rise to periodic switching behaviour are indicated with a filled square, showing an increase as the disk gets thinner. {\bf (b)} The FFT-power spectra, normalised with the amplitude of the 56~nm thick disk. {\bf (c)} The spatial distributions of the FTT-amplitude at a frequency corresponding to the breathing mode (thick disks, $h>35$~nm) or the radial spin waves modes $n=2$ and $n=3$ (thin disks, $h<35$~nm). For the 35~nm thick disk, no periodic switching behaviour occurs due to the overlap of both modes. The main peaks of the spectra are related to the $n=1$ radial mode.}
    \label{fig_SwitchingThick} 
\end{figure*}

In addition, our micromagnetic simulations reveal so-called hybrid modes, which consist of a superposition of two subharmonics. The switching frequency of the core polarisation alternates between $f_p=f/N$ and $f_p=f/(N+2)$, such that a hybrid mode can effectively be regarded as an \emph{even} subharmonic with mode number $N^*=N+1$, as illustrated in Fig.~\ref{fig_overview}(e) for $N^*=4$. However, these hybrid modes do not meet the requirement $r>0.9$ which we impose for periodic switching (e.g., $r=3/5$ if $N^*=4$) and the core polarisation averaged over time is nonzero, with the dominant polarisation being determined by the vortex' initial state. Nonetheless, the switching behaviour of hybrid modes sharply contrasts the incommensurate switching which is found when the amplitude and frequency of the external field are not tuned properly, as is clear from the example shown in Fig.~\ref{fig_overview}(f). This irregular behaviour encompasses both intermittent~\cite{PYL-13} and chaotic switching~\cite{PET-12, PYL-13, YOO-20}, which lie outside of the focus of this paper.\\ 

\subsection{Excitation mechanisms}
As a next step, we investigate how the switching diagrams are affected by the thickness of the disk. The disk thickness is increased from 21 to 56~nm in steps of 7~nm, by adding one additional layer at a time using the same simulation cell size as before. As shown in Fig.~\ref{fig_SwitchingThick}(a), the frequencies giving rise to periodic switching are decreasing for increasing thicknesses. To elucidate the origin of this effect, we identify the eigenfrequencies of each disk. An out-of-plane field of 30~mT is applied over the entire disk and turned-off after a period of 2~ns in order to excite spin waves~\footnote{The excitation of spin waves can also be achieved by applying a sinc pulse. It has been checked that the natural frequencies are independent of the pulse used (square or sinc) for the system we consider here.}. A fast Fourier transform (FFT) is performed on the $z$-component of the magnetisation, $m_z$, to transform the temporal oscillations of $m_z$ to the frequency domain. The eigenfrequencies can be determined from the FFT power spectra, which are shown in Fig.~\ref{fig_SwitchingThick}(b).\\

For the disks with a thickness of 21 and 28~nm (displayed at the right in Fig.~\ref{fig_SwitchingThick}), periodic switching behaviour is found at a frequency of about 15.8 and 14.8~GHz respectively. These are the eigenfrequencies of the radial spin wave (RSW) modes with mode number $n=2$ and $n=3$~\cite{BUE-05}, as evidenced by the FTT-amplitude spatial distribution diagrams shown in Fig.~\ref{fig_SwitchingThick}(c).\\

In contrast, the thicker disks (shown at the left in Fig.~\ref{fig_SwitchingThick}) do not exhibit periodic switching for field amplitudes below 100~mT when the frequency of the external field matches the RSW modes. Instead, the core reverses periodically by resonant excitation of the \emph{breathing mode}~\cite{KIM-14b}, which occurs at $f=11.2$~GHz, $f=9.7$~GHz, and $f=8.5$~GHz for a disk thickness of 42, 49, and 56~nm, respectively. Since the breathing mode is the periodic compression and expansion of the vortex core, the FTT-amplitude diagrams of Fig.~\ref{fig_SwitchingThick}(c) only display a peak around the central core region, with the amplitude of the outer region being zero.\\

Between these two regimes, there is a transition region where the RSW mode and the breathing mode overlap and no periodic switching behaviour occurs. This is the case for a 35~nm thick disk in the driving-field frequency range from 10 to 16~GHz, a part of which is shown in Fig.~\ref{fig_SwitchingThick}(a).\\ 

\subsection{Energy analysis}

\begin{figure}
\centering{
    \includegraphics[width=0.48\textwidth]{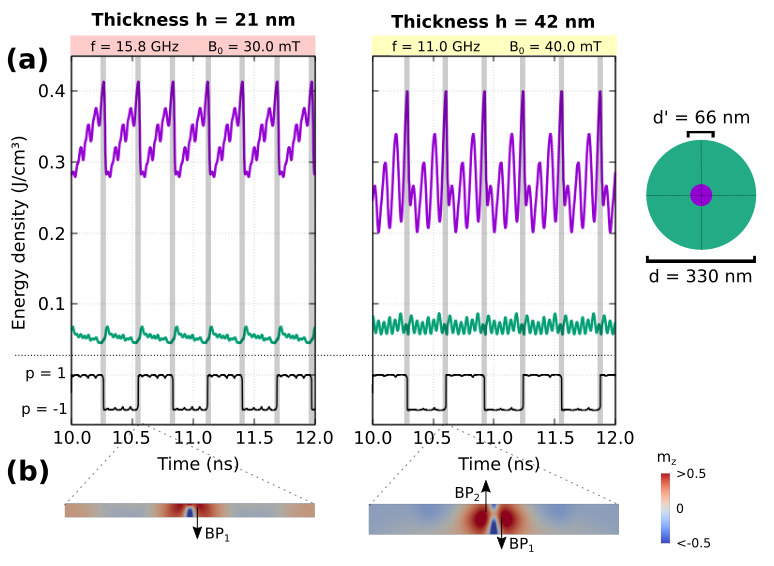}}
    \caption{{\bf (a)} The energy density of the central region (purple) and outer region (dark green) as a function of time, shown for a 21~nm thin disk (left panel) and a 42~nm thick disk (right panel). In the thin disk, spin waves transfer energy from the central to the outer region after a core reversal, as indicated by the gray bars. In the thick disk, the oscillating field $B_0 \sin (2 \pi f\ t) {\bf e}_z$ mainly pumps energy into the central region, without a transfer to the outer region, which is reminiscent of the breathing mode. {\bf (b)} An illustration of the switching mechanism governing the reversal of the core polarisation. The reversal occurs via the injection of either one Bloch point, labeled BP$_1$, at the surface (thin disk, left) or two Bloch points, labeled BP$_1$ and BP$_2$, in the middle (thick disk, right). The arrows indicate the direction in which the Bloch points move.}
    \label{fig_EnerThick} 
\end{figure}

To further elaborate on the difference between periodic switching via the breathing mode and the RSW modes, we analyse how the total energy density of the different regions evolves in time. For this analysis, the disk is subdivided into a central region with a diameter equal to $d'=66$~nm [purple in Fig.~\ref{fig_EnerThick}(a)] and an outer region encompassing the remaining 96\% of the disk [dark green in Fig.~\ref{fig_EnerThick}(a)]. We consider two cases. On the one hand a 21~nm thin disk on which we apply an external out-of-plane field with frequency $f=15.8$~GHz and amplitude $B_0=30.0$~mT in order to obtain a subharmonic switching mode $N=9$, shown on the left of Fig.~\ref{fig_EnerThick}(a). On the other hand a 42~nm thick disk, where the subharmonic switching mode $N=7$ is the result of a field frequency $f=11.0$~GHz and a field amplitude $B_0=40.0$~mT, shown on the right of Fig.~\ref{fig_EnerThick}(a).\\

For both cases, the energy density in the central region increases incrementally until a threshold is reached above which the formation of Bloch points (BPs) is energetically allowed, making it possible for the core polarisation to reverse. Notably, the energy threshold displays the same value for switching induced by a static out-of-plane field (the amplitude being more than 300~mT) and even for chaotic switching. For thin disks, the reversal occurs via the injection of one BP at the surface, after which the BP travels across the disk towards the other surface where it is expelled. The reversal in thick disks is governed by the injection of two BPs, which move in opposite direction and each disappear at a different surface. Both switching mechanisms are illustrated in Fig.~\ref{fig_EnerThick}(b). However, the switching mechanism is not related to which mode is excited (i.e., breathing or RSW). It is purely a result of the thickness, where we have verified that the difference is not due to the grid discretisation by using cells with a reduced $c_z$ for the 21~nm thin disk.\\

The fact that periodic switching occurs via excitation of the RSW mode in thin disks whereas it occurs via the breathing mode in thick disks, is evidenced by the time evolution of the energy density in the outer region. For the 21~nm thin disk, the energy density of the central region sharply drops after a reversal, where the released energy is transferred to the outer region in the form of spin waves, hence increasing the energy density of the outer region [see the left panel of Fig.~\ref{fig_EnerThick}(a)]. The energy density of the outer region then decreases steadily until the next switching event. This decrease contrasts the behaviour of the 42~nm thick disk, for which the energy density of the outer region fluctuates around a fixed value between two consecutive switching events, as shown in the right panel of Fig.~\ref{fig_EnerThick}(a). The energy of the oscillating external field is thus pumped mainly in the central region, without a transfer from the inner to the outer region. This is reminiscent of the breathing mode, since this mode is due to the compressing and expanding motion of the vortex core, i.e., the central region.

\subsection{Threshold field amplitude}
So far, we have mainly focused on the frequency of the external field. This showed that periodic switching takes place when the frequency matches the breathing mode or the RSW mode, depending on the disk thickness. For potential applications, the excitation amplitude should be minimal. We therefore shift our attention to the field amplitude needed to induce periodic switching behaviour, investigating the influence of {\it i)} the Gilbert-damping constant, {\it ii)} the disk size, and {\it iii)} the inhomogeneity of the applied field.    

\subsubsection{Gilbert-damping constant}
\begin{figure}
\centering{
    \includegraphics[width=0.48\textwidth]{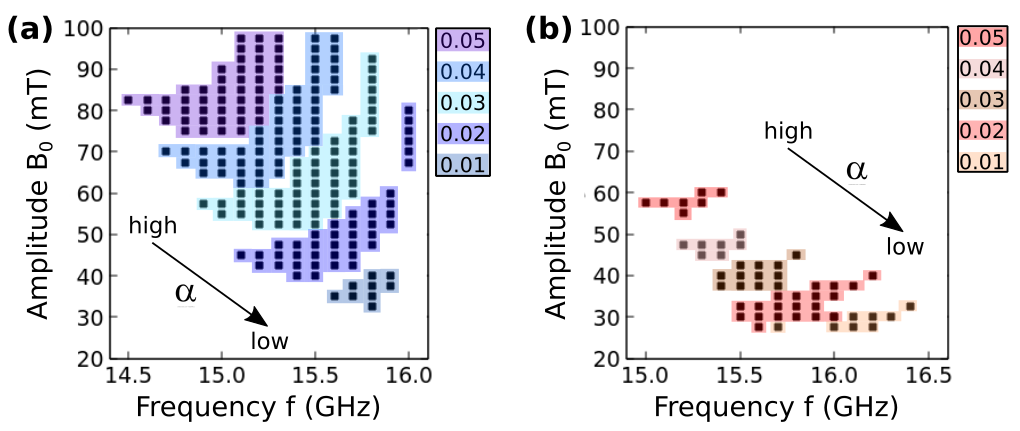}}
    \caption{The influence of the Gilbert-damping constant $\alpha$ on the regions with periodic switching, shown for the subharmonic switching mode {\bf (a)} $N=3$ and {\bf (b)} $N=9$. The results are obtained for a 21~nm thin disk with a diameter of 330~nm.}
    \label{fig_alpha} 
\end{figure}

The influence of the Gilbert-damping constant is investigated by acquiring the switching diagram of a 21~nm thin disk for different $\alpha$. Figure~\ref{fig_alpha}(a) and (b) display the ($f$,$B$)-points of the subharmonic switching mode $N=3$ and $N=9$, respectively. These results show that smaller damping gives rise to a lower threshold field, with the resonant frequencies being shifted towards slightly larger values, similar to a classical harmonic oscillator. The threshold field is lowered because the energy of the oscillating field is more effectively transferred to the disk due to the reduced energy dissipation (i.e., smaller $\alpha$). However, the regions of periodic behaviour are considerably smaller for $\alpha=0.01$ than for $\alpha=0.02$. The latter is therefore used throughout this work to trade-off between energy efficiency and robustness.  

\subsubsection{Disk size}

\begin{figure*}
\centering{
    \includegraphics[width=\textwidth]{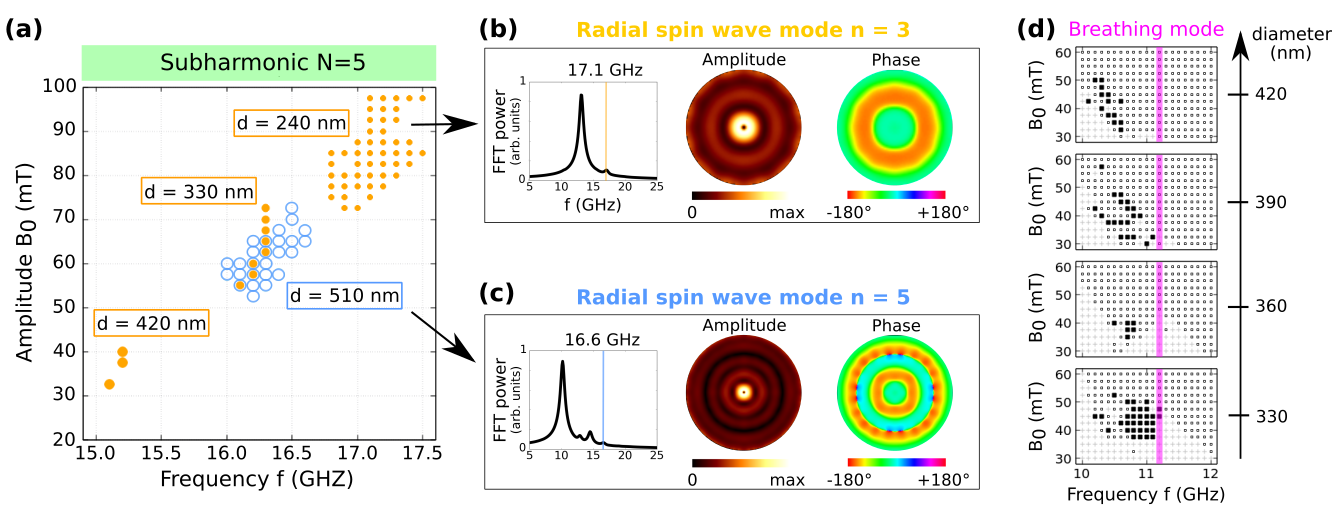}}
    \caption{{\bf (a)} The influence of the disk diameter on the ($f$,$B$)-points which give rise to periodic vortex core switching, for disks with a thickness of $h=21$~nm, illustrated for subharmonic switching mode $N=5$. {\bf (b)}-{\bf (c)} The FFT-spectrum of a 240~nm and 510~nm diameter disk. The spatial distributions of the FFT-amplitude and phase are shown at the eigenfrequency of the RSW mode $n=3$ and $n=5$, as indicated by the orange and blue line, respectively. {\bf (d)} For disks with a thickness of $h=42$~nm, periodic switching occurs at the eigenfrequency of the breathing mode of about 11~GHz, which remains largely unaffected when the diameter changes, as indicated by the pink line. The threshold field for periodic switching is about 40~mT, independent of the diameter.}
    \label{fig_diameter} 
\end{figure*}

The effect of the thickness on the threshold field amplitude was already shown in Fig.~\ref{fig_SwitchingThick}. When the core switches periodically due to resonant excitation of the breathing mode, lower field amplitudes are needed for thinner disks. A similar trend is observed for the RSW mode.\\

As a next step, we investigated the influence of the disk diameter, considering disks with a thickness of $h=21$~nm. Figure~\ref{fig_diameter}(a) shows the ($f$,$B$)-points which give rise to subharmonic switching mode $N=5$ for different diameters. The choice of $N=5$ is for illustrative purposes only, as the other subharmonics display the same qualitative trend.\\

A larger diameter lowers the required field amplitude as long as the periodic behaviour is caused by excitation of the same RSW mode. For instance, if the diameter is increased from $d=240$~nm to $d=420$~nm, the required field amplitude decreases from $B_0=72.5$~mT to $B_0=32.5$~mT, as depicted by the orange dots in Fig.~\ref{fig_diameter}(a). In addition, the frequency at which periodic switching occurs shifts towards slightly lower values, since a larger diameter reduces the eigenfrequency of the RSW modes~\cite{VOG-11}. In this case, the periodic switching is due to the excitation of the RSW mode $n=3$, which can be seen from the spatial distributions of the FFT-amplitude and phase of Fig.~\ref{fig_diameter}(b).\\    

However, decreasing the required field amplitude by increasing the disk diameter does not work indefinitely. For example, for a disk with a diameter of $d=510$~nm, no periodic switching occurs around the natural frequency of the $n=3$ RSW mode, neither at lower field amplitudes nor at higher. Instead, we find that the vortex core periodically reverses for frequencies close to the natural frequency of the $n=5$ RSW mode, the spatial distributions of which are shown in Fig.~\ref{fig_diameter}(c). The periodic switching happens at a field of more than 50~mT, as indicated by the blue dots in Fig.~\ref{fig_diameter}(a). Despite the larger diameter, the required field has not decreased compared to the $d=420$~nm disk. This indicates that the energy pumped into the outer edges of the system is not longer transferred efficiently to the core region, e.g., due to damping. As a consequence, the energy density of the core region does not reach the threshold value necessary for the formation of a Bloch point (see the left panel of Fig.~\ref{fig_EnerThick}). A minimal field must therefore be applied in order to induce (periodic) switching, which is about 30~mT for the material parameters outlined above.\\ 

A similar analysis can be made for thick disks, in which the periodic switching is related to the breathing mode. We focus on disks with a thickness of $h=42$~nm and a diameter ranging from 330 to 420~nm. The spectra of these disks reveal that the eigenfrequency of the breathing mode remains unaffected when the diameter changes. The regions of periodic switching are therefore excepted to occur near a frequency of 11.2~GHz (see Fig.~\ref{fig_SwitchingThick}). The results shown in Fig.~\ref{fig_diameter}(d) corroborate this picture: periodic switching occurs at frequencies between 10.5 and 11.0~GHz for all disk diameters. During the excitation of the breathing mode, the energy of the oscillating external field is mainly pumped in the central region. Because this central region is independent of the total disk diameter, the threshold field of periodic switching is almost constant, being 40~mT for all considered diameters. This is in sharp contrast to thin disks, where the required field halves by increasing the diameter from 330 to 420~nm [see Fig.~\ref{fig_diameter}(a)].  

\subsubsection{Inhomogeneous fields}
So far, the field has been applied homogeneously over the entire disk. Additional simulations show that periodic switching also emerges when the field is only applied in the central region around the core. Compared to the homogeneous case, however, the threshold field to induce periodic core reversals in a 21~nm thin disk with diameter $d=330$~nm and central region $d'=66$~nm increases from 30 to 40~mT (subharmonic switching mode $N=9$) and from 40 to 50~mT (subharmonic switching mode $N=3$). This larger field amplitude is needed to compensate for the energy transfer from the outer to the central region.\\

In a 42~nm thick disk, the switching is due to excitation of the breathing mode, for which only the energy density of the central region is relevant. This means that the minimal field amplitude should remain constant, irrespective of whether the field is applied over the entire disk or over the central region only. This hypothesis is confirmed by our simulations, yielding a threshold field of about 40~mT for all cases. 

\section{Conclusion}
We show that the polarisation of the vortex core can be reversed periodically in magnetic nanodisks by applying an oscillating out-of-plane field. Two different regime could be identified. The frequency of the field should be tuned to the eigenfrequency of the radial spin wave modes in the case of thin disks ($<35$~nm) or the eigenfrequency of the breathing mode in the case of thick disks ($>35$~nm). For thin disks, the threshold field amplitude required for periodic switching can be lowered to about 30~mT by increasing the disk diameter. For thick disks, in contrast, the minimal field is largely unaffected by the disk diameter, as only the energy density of a central region around the vortex core is relevant to excite the breathing mode.
Our results contribute to the understanding of the switching mechanisms in magnetic nanodisks, which are of technological interest due to their potential in non-volatile memory devices.

\begin{acknowledgments}
This  work  was  supported  by  the  Fonds  Wetenschappelijk  Onderzoek  (FWO-Vlaanderen) with senior postdoctoral research fellowship 12W7622N (J.L.).\\


\end{acknowledgments}

\bibliographystyle{unsrtnat}
\bibliography{biblio.bib}

\end{document}